\documentclass[aps,prl,preprint,showpacs,groupedaddress]{revtex4}

\usepackage{graphicx}
\usepackage[latin1]{inputenc}

\begin{document}
\title{Experimental growth law for bubbles in a "wet" 3D liquid foam}


\author{J\'er\^ome Lambert}
\affiliation{
G.M.C.M. Universit\'e Rennes 1 - UMR CNRS 6626 
B\^atiment 11A - Campus Beaulieu 35042
Rennes cedex France}
\author{Rajmund Mokso}
\affiliation{
Imaging Group - ESRF, BP 220, 38043 Grenoble Cedex 9, France}
\author{Isabelle Cantat}
\affiliation{
G.M.C.M. Universit\'e Rennes 1 - UMR CNRS 6626 
B\^atiment 11A - Campus Beaulieu 35042
Rennes cedex France}
\author{Peter Cloetens}
\affiliation{
Imaging Group - ESRF, BP 220, 38043 Grenoble Cedex 9, France}
\author{Renaud Delannay}
\affiliation{
G.M.C.M. Universit\'e Rennes 1 - UMR CNRS 6626 
B\^atiment 11A - Campus Beaulieu 35042
Rennes cedex France}
\author{James A. Glazier}
\affiliation{
Department of Physics, Swain Hall West 159, Indiana University, 727 East Third
Street, Bloomington, IN 47405-7105, U.S.A.}
\author{Fran\c cois Graner}
\affiliation{
LSP - UMR 5588 - 140 Avenue de la physique, BP 87 
- 38402 Saint Martin d'H\`eres, France}

\begin{abstract} 
We used X-ray tomography to characterize the geometry of all bubbles in a liquid foam of average liquid fraction $\phi_l\approx 17\,\%$ and to follow their evolution, measuring the normalized growth rate $\mathcal{G}=V^{-\frac{1}{3}}\frac{dV}
{dt}$ for 7000 bubbles. While $\mathcal{G}$ does not depend only on the number of faces of a bubble, its average over $f-$faced bubbles scales as $G_f\sim f-f_0$ for large $f$s at all times. We discuss the dispersion of $\mathcal{G}$ and the influence of $V$ on $\mathcal{G}$.
\end{abstract}
\maketitle


Liquid foams consist of bubbles of gas 
separated by a continuous liquid phase  
occupying a fraction $\phi_l$ of the foam's 
volume. Liquid foams coarsen because gas slowly 
diffuses through the liquid films from high pressure bubbles to 
low pressure bubbles so the high pressure bubbles eventually disappear. The dynamics of the coarsening differs in the two limiting cases of dry and infinitely wet foams.\\

In a bubbly liquid, with $\phi_l \sim 1$, bubbles 
are spherical and well separated.
As in emulsions and diphasic materials, the bubbles coarsen via 
 "Oswald ripening" as theoretically  described by Lifshitz, Slyozov and Wagner (LSW) 
\cite{lsw, wagner61} using a mean-field approach.\\

In the dry foam limit,  $\phi_l  \ll 1$, bubbles 
are polyhedral and touch each other. 
The gas flux between two neighboring bubbles is proportional to their pressure difference, hence to the mean curvature of the film separating them. 
From simple dimensional arguments \cite{glazier93}, the growth rate of an individual bubble of volume $V$ in a 3D foam must be of the form:
\begin{equation}
\frac{\textrm{d}V}{\textrm{d}t}= V^{\frac{1}{3}}\mathcal{G}
\end{equation}
$\mathcal{G}$ depends both on the shape of the bubble rescaled to unit volume and on physico-chemical characteristics of the liquid, gas and of the surfactant through the effective gas diffusivity $D_{eff}$:\\
\begin{equation}\label{equation2}
\mathcal{G}= - D_{eff}\, \int_S \frac{H\textrm{d}S}{V^{\frac{1}{3}}}\,,
\end{equation}
where $H$ is the mean curvature of a surface element d$S$ of the bubble.
Theoretical treatment of coarsening in dry foams usually make an analogy with 2D foams, for which von Neumann's law (vNM), which is exact, states that the growth rate of the area of a bubble is proportional to $n-6$, where $n$ is its number of edges. This law has been validated experimentally. In 3D, in contrast, the existence of the extra radius of curvature means that the growth rate of a bubble does not depend only on its number of faces \cite{jurine,glazier00}. Thus deriving a growth law in 3D requires additional assumptions and results are only approximate. 
Attempts to relate the growth rates of individual bubbles or the average growth rates of classes of bubbles to simple geometrical properties have had a mixed success.
Indeed many attempts have been done to find a relation between the right-hand-side (rhs) of eq. \ref{equation2} and simple geometrical properties. Mullins \cite{mullins89} and several groups (see \cite{jurine} and references therein) derived estimates for this rhs based on ideal regular bubbles and obtained $\mathcal{G}_{M}\sim f^{1/2}$ for large $f$. 
Numerical simulations of coarsening have computed $\mathcal{G}$ for large numbers of bubbles.
Potts Model \cite{glazier93, thomas06}, vertex model \cite{fuchizaki95} and Surface Evolver \cite{wakai00} simulations all exhibit a strong correlation between bubbles' number of faces and growth rate with a small dispersion. We can then check if Mullins' model holds on average for realistic foams by evaluating the average values $\mathcal{G}_f=<\mathcal{G}>_f$ over sets of $f-$faced bubbles. The data admit both $G_f\sim (f-f_0)$ and $G_f\sim (f-f_0)^{\frac{1}{2}}$ fits.
Experimental measurements of $G_f$ using optical tomography \cite{monnereau98b,monnereau98} and NMR \cite{gonatas95,prause95,prause99,theseprause}
observed too few bubbles to distinguish $(f-f_0)$ from $(f-f_0)^{\frac{1}{2}}$.

Here we investigate foam 
coarsening experimentally in the previously unstudied intermediate wetness regime with $\phi_l \sim 
10-20\,\%$, which is not known in theory nor 
in simulations, but which is widely used in most foam applications \cite{weaire}. 
We determine $\mathcal{G}$ and $G_f$ for several thousand bubbles and study the cross-over between the LSW 
and the vNM regimes.
Recent progress in X-ray tomography 
\cite{lambert2005} permits us to take 3D images of thousands 
of bubbles in 2 minutes, and allows us to monitor a foam 
for many hours.
We find that, for a given set of $f-$faced 
bubbles, the dispersion of $\mathcal{G}$ is small.
Therefore we focus on $G_f$. While the shape of the distribution of the number 
of faces per bubble changes during the 
experiment (implying that our foam has not reached a statistically invariant scaling state), the shape of $G_f$ vs.~$f$ stays constant. This quantity thus seems to be a 
robust parameter to characterize foam coarsening.
$G_f$ depends linearly on $f$ for large f. 
We then study the dispersion of $\mathcal{G}$ as a function of $f$ and $V$ and we finally discuss whether $f$ or $V$ is more 
relevant to determine the dynamics of a bubble, on the basis of the strong correlation existing between both quantities through Lewis' law.\\

The X-ray tomography apparatus at the ID19
beamline of the European Synchrotron Radiation facility (ESRF) Grenoble, France, enables us to visualize the coarsening of a wet liquid foam contained in a $1\,$cm$^3$ cubic volume. The spatial
resolution of the images is $10\,\mu$m in each direction, smaller than the smallest
bubbles which appear in at least two successive images (i.e. any bubble smaller than $10\,\mu$m diameter disappears in less than 2 minutes. It thus never enters our calculations of growth rates).
 Thus we can image all bubbles which we can analyze (by definition, we do not include in $G_f$ bubbles which disappear within one time step). Obtaining adequate image contrast
between liquid and gas phases is a major difficulty in this kind of experiment since
the liquid phase occupies only a small percentage
of the total volume - the liquid fraction $\phi_{l}$ \footnote{To achieve a relatively high contrast, we imaged wet
foams with a beam of energy $15\,$keV.}. 
In addition, the bubble walls must not move significantly during the time required to acquire an image otherwise topographic reconstruction of the foam will fail. The characteristic evolution time for gas diffusion in $\sim1\,$mm$^3$ air bubbles is several minutes. Our experimental acquisition
time of $2\,$min per image enabled us to clearly identify and follow bubbles during
their evolution. To reduce the complicating effects of gravitational drainage of the foam and to maintain a constant and relatively high liquid fraction, we used a high-viscosity liquid composed of $100\,$mL
de-ionized water, $0.1\,$g sodium dodecyl phosphate, $0.003\,$g dodecanol and $1\,$g
gelatine (not enough to gelify the liquid). In addition, we used a pump to provide a constant supply of liquid to the top of the foam which maintained the liquid fraction, decreasing from $20\%$ at the top
to $10\%$ at the bottom \cite{lambert2005}.\\
We identified bubbles in an image using a watershed  
method detailed in \cite{lambert2005}.
We measured the liquid fraction in a region by summing
the number of pixels of that volume containing more than a threshold amount of liquid. $\phi_l$ was thus estimated with an error $\Delta\phi_l=0.5\%$.


We developed a correlation algorithm using the commercial \textit{Aphelion} \footnote{URL: http://www.adcis.net}
software package to track
bubbles from image to image, enabling us to determine their growth rates.\\

Determining a bubble's number of faces in a wet foam is ambiguously defined
\cite{lambert2005}, since neighboring bubbles may be distant. We compared the
distance separating two bubbles to the characteristic size of their vertices, i.e. the crossing of the bubbles' edges where most of the liquid accumulates. We assumed that if the distance between two
bubbles
was less
than half this characteristic size, they were neighbors and shared a face. Thus small bubbles located at vertices may have fewer than 4 faces. We automatically discarded from our analysis bubbles which touched the container walls.\\  

\begin{figure}[h]
 \includegraphics[width=8cm]{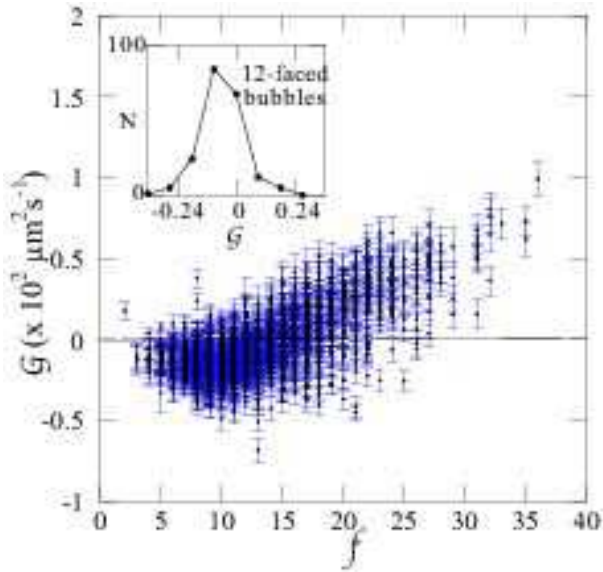}
 \caption{\label{fig1} Scatterplot of $\mathcal{G}$ vs.~$f$ in the topmost slice ($\phi_l=20\,\%$) of the experimental cell. Inset: histogram of $\mathcal{G}$ for 12-faced bubbles.}
 \end{figure}

48 minutes after production, the foam's liquid fraction varied from 20\% at its top to 10\%
at its bottom as a result of the continuous liquid injection at its top.
To test the influence of the liquid fraction on foam dynamics, we defined four subregions corresponding to different heights and thus wetnesses in our column. \\

for each image, we measured each bubble's number of neighbors $f$ and growth rate $\mathcal{G}$.
figure \ref{fig1} shows $\mathcal{G}$ vs.~$f$ for the 3582 bubbles in the topmost horizontal slice of the cell $48 \,$min after foam creation, and the distribution of $\mathcal{G}$s for $f=12$ (inset). Error bars on the measurement of $\mathcal{G}$ result of the incertitude on the measurement of $V$ at two successive times. $\Delta V$ was determined for each bubble by assuming that a 1 voxel thick sheet could randomly be added or removed from each of its faces. 

\begin{figure}
 \includegraphics[width=8cm]{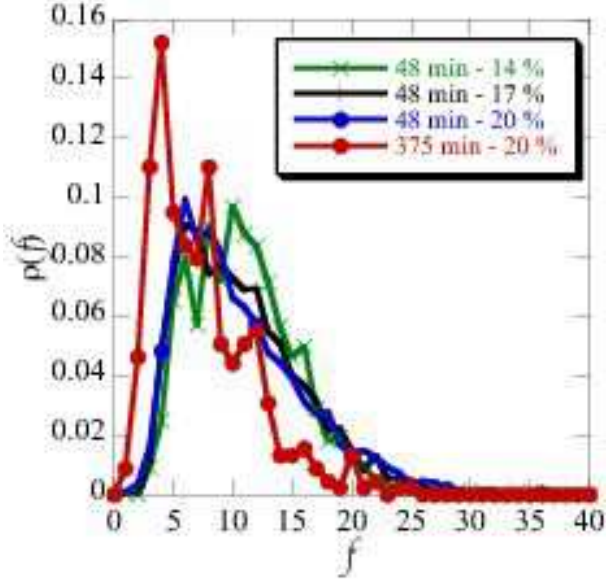}
 \caption{\label{fig2}Solid red and blue lines: $\rho(f)$ for $\phi_l=20\%$, $48$ min and $375$ min after foam creation. Solid green and black lines: $\rho(f)$ for $\phi_l=14\%$ and $17 \%$, $48$ min after foam creation.}
 \end{figure}

\begin{figure}
 \includegraphics[width=8cm]{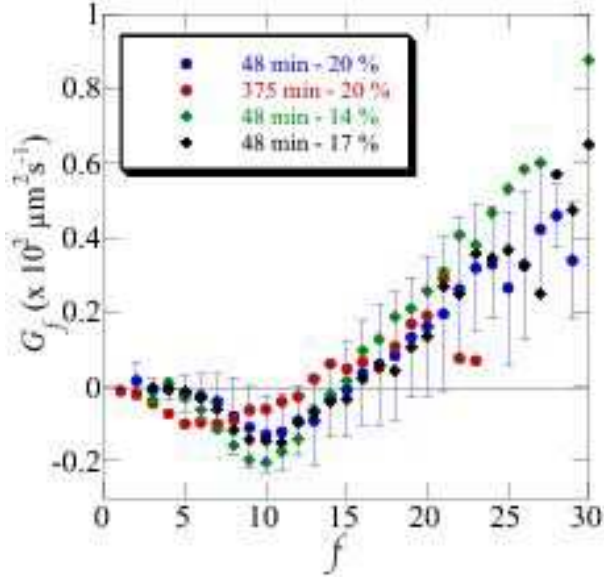}
 \caption{\label{fig3}Blue and red points $G_f$ vs.~$f$ at times  $48\,$min and $375\,$min after the making of the foam for $\phi_l=20\,\%$. The standard deviation of $\mathcal{G}$ around $G_f$ for each $f$ is pictured by the bars.
  Green and black points : $G_f$ vs.~$f$ at time $48\,$min for $\phi_l=14\,\%$ and $\phi_l=17\,\%$.}
 \end{figure}

$\mathcal{G}$ varies for different bubbles with the same $f$s to a much greater extent than the experimental uncertainty. Hence it is not an exact function of the unique variable $f$, although the two quantities are correlated, showing that 3D bubbles do not obey an exact equivalent of von Neumann's law.
However for the ensemble of bubbles with a given $f$, the dispersion of $\mathcal{G}$ around its average is not large, as the inset of fig. \ref{fig1} shows for f=12, since one can clearly see the increase of $\mathcal{G}$ with $f$: The dispersion $\sigma_\mathcal{G}$ coincides with the variation of $G_f$ for $\Delta f\approx5$ (with $f >10$). We therefore compute the average value $G_f$ to investigate its dependence on $\phi_l$ and the structure of the foam.\\
Foam structure changes during coarsening.
We characterize the structure by measuring the distribution of bubbles' number of faces $\rho(f)$. figure \ref{fig2} shows $\rho(f)$ at different times and for different values of $\phi_l$.
The shape of the distribution varies significantly with time and liquid fraction.
 The average number of neighbors ranges from $10.7\, \pm\,5.5 $ ($\phi_l=20\%$) to $11.0\,\pm\,4.5$ ($\phi_l=14\%$). Figure \ref{fig2} also shows that $\rho(f)$ varies significantly from $48$ min to $375$ min after foam creation, hence the structure is still far from any scale invariant regime in which $\rho(f)$ would be constant \footnote{In a scaling state, a characteristic length $l(t)$ keeps growing with time, while all the distributions and correlations, approprietely rescaled by powers of $l(t)$ when necessary, are constant.}. \\
Figure \ref{fig3} shows $G_f$ at two different times for different values of $\phi_l$. Despite a shift in the value of $f$ at which $G_f$ crosses 0, the general dependence of $G_f$ vs.~$f$ is unchanged:  $G_f$ goes to $0$ as $f\,\rightarrow \,0$ for $f<10$, and $G_f$ is linear for $f>10$.\\

\begin{figure}
 \includegraphics[width=8cm]{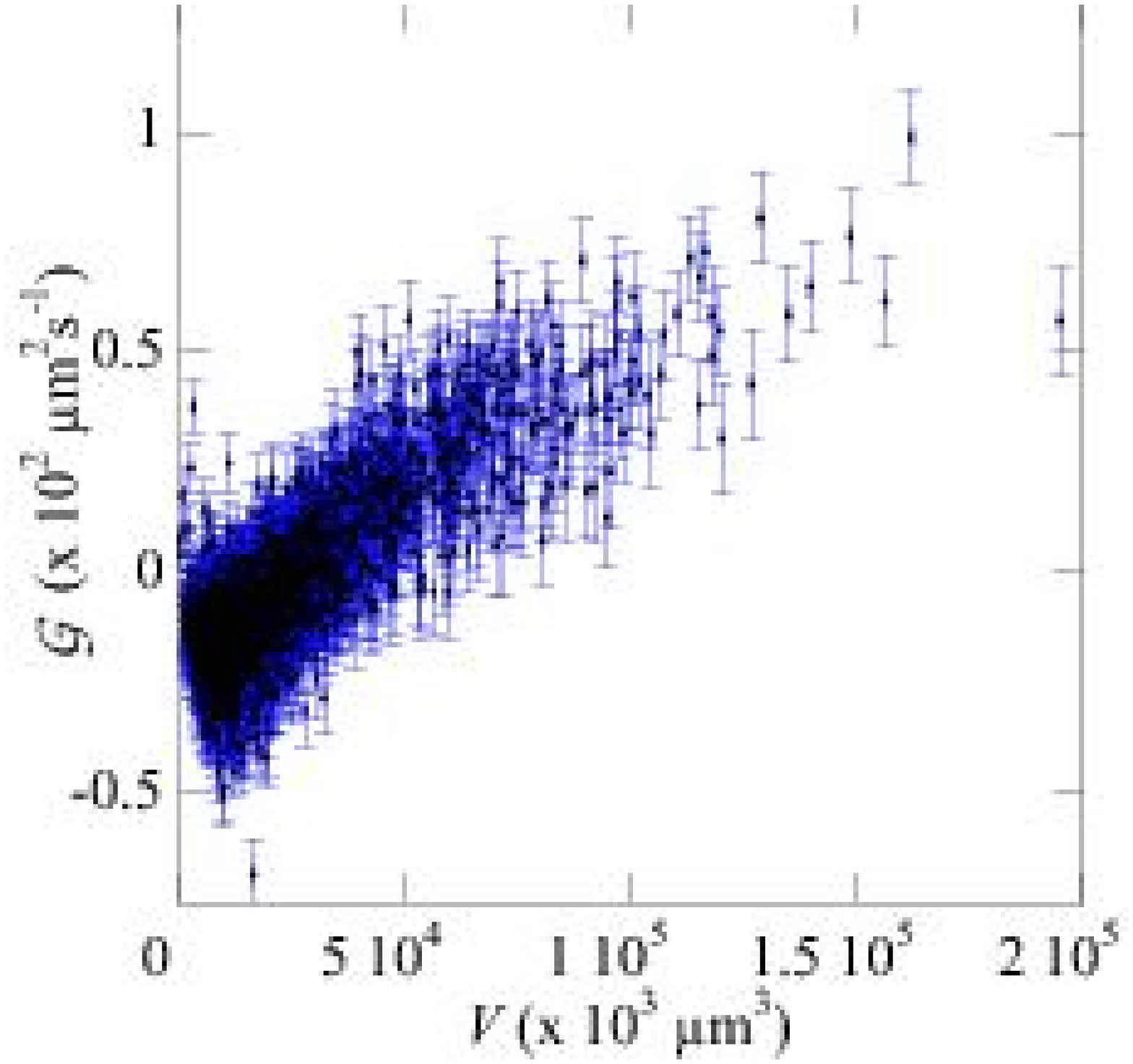}
 \caption{\label{fig5}Growth rate $\mathcal{G}$ vs.~$V$, volume of the bubbles inside the topmost slice ($\phi_l=20\,\%$) in the $48\,min$ old foam.}
 \end{figure}

Since in the limit $\phi_l\rightarrow 1$ $\mathcal{G}$ should depend on $V$ only , we plot $\mathcal{G}$ vs.~$V$ in Figure \ref{fig5}. As expected, both quantities correlate.

\begin{figure}
 \includegraphics[width=9cm]{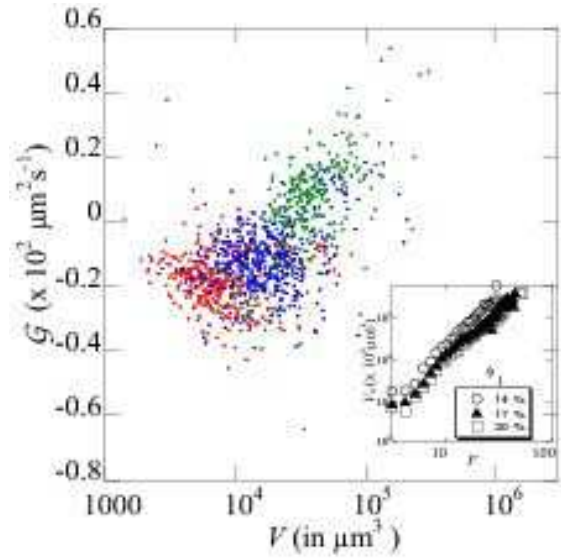}
 \caption{\label{fig4} Scatterplot of the growth rate $\mathcal{G}$ vs.~$V$ for 8 (red),12 (blue) and 16-faced (green) bubbles. Inset: $<V>_f$ as a function of $f$ for $\phi_l=14,\,17$ and $20\%$.}
 \end{figure}

In order to check if the number of faces describes better a bubble's behavior than its volume as it is expected in the dry foam limit \cite{glazier93}, we scatter-plot $\mathcal{G}$ vs.~$V$ for three different sets of $f-$faced bubbles in fig. \ref{fig4}. It is clear that, at fixed $f$, $\mathcal{G}$ increases slightly with $V$. The shift of the clouds corresponding to different $f-$set over the $V$ axis can be explained by the strong correlation between $<V>_f$ and $f$, that we plot in the inset of fig. \ref{fig4}, usually known as Lewis' law \cite{thomas06, wakai00, weaire}. 

We experimentally observe a power law relation between $<V>_f$ and $f$ : $V_f=<V>_f\sim f^\alpha$, with $\alpha\approx 2.2$ here. However this law does not account for the slight increase of $\mathcal{G}$ vs.~$V$ for $f=16$. \\  
 
We compared the dispersion of the data points for scatter plots of $\mathcal{G}$ vs.~$f$ (fig. \ref{fig1}) and vs.~$V$ (fig.~ \ref{fig5}). for $f=12$ (476 bubbles) we obtained a standard deviation for $\mathcal{G}$ of $\Delta \mathcal{G}=10.5\,\mu$m$^2.$s$^{-1}$. On the basis of the empirical Lewis' law $V_f\sim f^{2.2}$ we plotted on fig.~ \ref{fig4}, we compute a $\Delta V$ corresponding to $\Delta f=1$ and find $\Delta \mathcal{G}=12.1\,\mu$m$^2.$s$^{-1}$ around $V_{12}$.
This small difference does not allow to conclude that one parameter is better than the other.
Future experiments using foams of very different structures (bidisperse, monodisperse, scaling state...) and liquid fractions may allow to obtain various relations between $f$ and $V$, and thus to determine which parameter, $V$ or $f$, influences $\mathcal{G}$ most significantly.\\

%
%
Our experiments studied a moderately wet foam ($14\%<\phi_l<20\%$), for the dynamics of which we have no theoretical predictions. $\mathcal{G}$ does not depend uniquely on $f$ and $G_f$ disagrees with mean-field theories for both wet and dry foams. $G_f$ increases linearly with $f$ for $f>10$ as Glazier numerically observed using Potts' simulations \cite{glazier93}. Thomas \textit{et al} \cite{thomas06} have confirmed Glazier's results and extended them to additional initial bubble size distributions. 
As in dry foam simulations, $G_f$ increases to zero as $f$ decreases for small $f$s \cite{glazier93, thomas06}.  Let us notice that there are still some spurious bubbles due to the noise in image segmentation. Some of them are artificially correlated, but their number has been evaluated to be very small, and their influence on the results is poor as was checked by introducing a filter on the size of the bubbles taken into account.

In contrast to 2D, the dispersion of $\mathcal{G}$ around $G_f$ is intrinsic to the foam, as $f$ does not suffice to describe the dynamics of a bubble. 
In addition we have checked the stability of the linearity of $G_f$ for early stages of the evolution of the foam. The foams contains too few bubbles at late stages to compute statistically relevant $G_f$s. It will be of great interest to observe $G_f$ in later stages, in particular during a possible scaling regime. 

\bibliography{biblio_mousses,bib}
\end{document}